\documentclass[12pt,oneside,openany]{article}
\usepackage{amsfonts,amssymb,graphicx}
\usepackage{cite}
\usepackage[affil-it]{authblk}
\usepackage{indentfirst}
\usepackage{newlfont}
\usepackage{amsthm}
\usepackage{amsmath}
\usepackage{latexsym}
\usepackage{eucal}
\usepackage{eufrak}
\usepackage{amssymb}
\usepackage{mathrsfs}
\usepackage{graphicx}

\begin{document}
\title{Inverse problem robustness for multi-species mean field spin models}

\author[1]{Micaela Fedele}
\affil[1]{
Courant Institute of Mathematical Sciences, New York University}
\author[2]{Cecilia Vernia}
\affil[2]{Dipartimento di Scienze Fisiche Informatiche e Matematiche, Universit\`a di Modena e Reggio Emilia}
\author[3]{Pierluigi Contucci}
\affil[3]{Dipartimento di Matematica, Universit\`a di Bologna}
\date{}

\maketitle

\begin{abstract}
The inverse problem method is tested for a class of mean field statistical
mechanics models representing a mixture of particles of different species. 
The robustness of the inversion is investigated for different values
of the physical parameters, system sizes and independent samples. We show how
to reconstruct the parameter values with a precision of a few percentages.
\end{abstract}

\section{Introduction}
In recent years there has been an increasing interest in studying the inverse problem
in statistical mechanics mostly due to the fact that the thermodynamic formalism on a macroscopic 
base has proved to be effective in a variety of scientific applications that span from the investigation
of real neural networks in biology \cite{MC,SM,RTH} to behavioral ethology for flocks \cite{BCGMSVW}. 
In this paper we are interested in a particular class of models \cite{CG,GC}
that have naturally emerged within the application of the statistical mechanics formalism
to socio-economic sciences (see also \cite{MB,GGS} and references therein). Their first and most 
elementary appearance can be traced back 
to the so called discrete choice theory
proposed by Daniel Mc Fadden \cite{McF} after his celebrated success in predicting 
the number and distribution of customers of the Bay Area Rapid Transit before its construction.
Discrete choice theory doesn't contain interaction between individuals and from the statistical
mechanics point of view can be seen as a mixture of a finite number of discrete perfect gases;
its inverse problem is mathematically elementary and its efficiency amounts to the proper
identification of the different species of particles (see also \cite{Fa}). 
The necessity to include the interaction among agents
led W. Brock and S.N. Durlauf \cite{BD} to introduce, within the socio-economic context,
the simplest interaction structure which is given by the mean field Curie-Weiss Hamiltonian
model. In order to successfully generalize the discrete choice to the interacting case,
it was defined in \cite{CG} a multi-species mean field model.
In this paper we propose a robustness test of the inverse problem in the multi-species mean field case. 
We start from the knowledge of the exact solution of the model, both in the single populated system and in the
bi-populated one, not only at the thermodynamic limit in analytic form but also at finite and increasing number
of particles by accurate numerical approximations. This, together with the standard criterion of maximum 
likelihood, provides a relation between experimental and theoretical quantities and allows
to tackle the computation of the free parameters of the model, namely interactions and magnetic fields, 
from observed data. 

We then generate the equilibrium configuration of the models, at different system sizes
and for different values of the parameters i.e. interaction strength and magnetic field. 
By use of the inversion formulas we show how the reconstruction of the parameters is
achieved and how his robustness depends on both system size and number of independent 
samples used. \

The paper is organized as follows. In section 2 we recall briefly the Curie-Weiss model and we review how to 
solve the inverse problem in this single-population model. The generalization of such a model to 
systems composed of many interacting groups (the multi-species mean field model) and the solution of the corresponding inverse problem are presented in section 3. Section 4 presents and discusses a set of numerical tests for both 
the single and the bi-populated system for finite number of particles and finite number of samples. We first
investigate how the average quantities, magnetization and susceptibility behave for increasing system sizes
in the standard Curie Weiss model. We find, in particular, that while the magnetization is monotonically increasing, in agreement with the first Griffiths
Kelly Sherman inequality \cite{Gr1,Gr2,KS}, the susceptibility has a monotonicity direction that changes with the values
of the coupling constant with respect to its critical point. Both quantities reach their limiting value at the speed
of the inverse volume. We then investigate how the experimental magnetization and susceptibility at fixed volume 
depend on the number of samples and stabilize when their number increases. The effectiveness of the inversion
is tested for different values of the coupling constants and magnetic fields. The same procedure is applied to
the bi-populated model and again the robustness of the inversion is tested for different values of the parameters.
We find in all cases that the inverse method reconstructs, with a modest amount of samples, the values of the 
parameters with a precision of a few percentages.

\section{Inverse problem for the Curie-Weiss model}\label{CWSec}

Denoting with $N$ the size of the population, the Curie-Weiss model is defined by the Hamiltonian:
\begin{equation}\label{Hami.curie}
H_{N}(\boldsymbol{\sigma})=-\frac{J}{2N}\sum_{i,j=1}^{N}\sigma_{i}\sigma_{j}-h\sum_{i=1}^{N}\sigma_{i}
\end{equation}
where $\sigma_{i}\in\{\pm 1\}$ is the spin of the particle $i$ (individual), the parameter $J>0$ is the coupling constant and $h$ is the value of the magnetic field. The joint probability of a configuration of spins $\boldsymbol{\sigma}=(\sigma_{1},\dots,\sigma_{N})$ is given by the  Boltzmann-Gibbs measure:
\begin{align}\label{BG.curie}
P_{N, J, h}\{\boldsymbol{\sigma}\}&=\dfrac{\exp(-H_{N}(\boldsymbol{\sigma}))}{\sum\limits_{\boldsymbol{\sigma}\in\Omega_{N}}\exp(-H_{N}(\boldsymbol{\sigma}))}\nonumber\\
&=\dfrac{\exp\bigg(N\Big(\frac{J}{2}m_{N}^{2}(\boldsymbol{\sigma})+hm_{N}(\boldsymbol{\sigma})\Big)\bigg)}{\sum\limits_{\boldsymbol{\sigma}\in\Omega_{N}}\exp\bigg(N\Big(\frac{J}{2}m_{N}^{2}(\boldsymbol{\sigma})+hm_{N}(\boldsymbol{\sigma})\Big)\bigg)}
\end{align}
where $\Omega_{N}=\{-1,1\}^{N}$ and $$m_{N}(\boldsymbol{\sigma})=\frac{1}{N}\sum_{i=1}^{N}\sigma_{i}$$ is the magnetization of the configuration $\boldsymbol{\sigma}$. We point out that, in the inverse problem, the usual inverse temperature parameter $\beta$ is absorbed within the two free parameters $J$ and $h$.

We will denote by $\omega (\cdot)$ the expectation value with respect to $P_{N, J, h}$.
Heuristically, this distribution favors both the agreement of people's
choices or opinions $\sigma_i$, with an external influence $h$, and the
agreement between individuals, being $J$ positive (whereas  for $J<0$ it would favor disagreement). 

The inverse problem amounts to compute the values of $J$ and $h$ starting from the knowledge of the magnetization
average and variance. Of course, when dealing with real phenomenological data its solution is made in two steps.
The first is the identification of the analytical inverse formula providing a possible explicit expression of the
free parameters ($J$ and $h$) in terms of the mentioned macroscopic thermodynamic variables. The second is
the evaluation with statistical methods of the macroscopic variables starting from real data. The problem is 
generically well posed because the unknown parameters (interaction and magnetic field) are as many as the measured phenomenological quantities (average magnetization and its fluctuation).\\
Let start by observing that when $h\neq 0$ and $J>0$ or $h=0$ and $J<1$, the Curie-Weiss model satisfies the following property (see \cite{E})
\begin{equation}\label{identita.curie}
\lim_{N\rightarrow\infty}\omega(m_{N}(\boldsymbol{\sigma}))=m(J,h)
\end{equation}
where $m(J,h)$ is the stable solution of the model mean-field equation
\begin{equation}\label{campo.medio.curie}
m(J,h)=\tanh(Jm(J,h)+h).
\end{equation} 
By differentiating the identity (\ref{identita.curie}) with respect to $h$ we obtain:
\begin{equation*}
\lim_{N\rightarrow\infty}\frac{\partial}{\partial h}\omega(m_{N}(\boldsymbol{\sigma}))=\chi
\end{equation*}
where
\begin{equation}\label{chi.2}
 \chi=\dfrac{\partial m(J,h)}{\partial h}=\dfrac{1-m^{2}(J,h)}{1-J(1-m^{2}(J,h))}
\end{equation}
and
\begin{align}\label{chi.1}
\frac{\partial}{\partial h}\omega(m_{N}(\boldsymbol{\sigma})) &=\frac{\partial}{\partial h}\bigg(\dfrac{\sum_{\boldsymbol{\sigma}\in\Omega_{N}}m_{N}(\boldsymbol{\sigma})\exp(-H_{N}(\boldsymbol{\sigma}))}{\sum_{\boldsymbol{\sigma}\in\Omega_{N}}\exp(- H_{N}(\boldsymbol{\sigma}))}\bigg)\nonumber\\
&=N\Big(\omega(m_{N}^{2}(\boldsymbol{\sigma}))-\omega^{2}(m_{N}(\boldsymbol{\sigma}))\Big).
\end{align}
In particular, the right hand side of the last identity defines the finite size susceptibility $\chi_{N}$.
By putting together (\ref{chi.2}) and (\ref{chi.1}) we can compute the parameter $J$ from the average value and the variance of the magnetization in the thermodynamic limit:
\begin{equation}\label{J.inverso.curie}
 J=\frac{1}{1-\lim\limits_{N\rightarrow\infty}\omega^2( m_{N}(\boldsymbol{\sigma}))}-\frac{1}{\lim\limits_{N\rightarrow\infty}N\Big(\omega( m_{N}^{2}(\boldsymbol{\sigma}))-\omega^{2}( m_{N}(\boldsymbol{\sigma}))\Big)}.
\end{equation}
The external field $h$ is obtained, in the large volume limit, by inverting the mean-field equation (\ref{campo.medio.curie})
\begin{equation}\label{h.inverso.curie}
h=\tanh^{-1}\left(\lim_{N\rightarrow\infty}\omega( m_{N}(\boldsymbol{\sigma}))\right)-J\lim_{N\rightarrow\infty}\omega( m_{N}(\boldsymbol{\sigma}))
\end{equation}
where $J$ is given by (\ref{J.inverso.curie}).
Formulas (\ref{J.inverso.curie}) and (\ref{h.inverso.curie}) solve the analytical inverse problem for the Curie-Weiss model as $h\neq 0$ and $J>0$ or $h=0$ and $J<1$. 
On the other hand, if $h=0$ and $J>1$ the equation (\ref{campo.medio.curie}) has two different stable solutions $\pm m(J,0)$ (see \cite{E}). Thus, the identity (\ref{identita.curie}) is not verified. In this case the inverse problem can be solved, by observing that there exists $\epsilon>0$ such that, whenever $m_{N}(\boldsymbol{\sigma})\in(\pm m(J,0)-\epsilon,\pm m(J,0)+\epsilon)$, the following limit holds (see \cite{ENR})
\begin{equation}\label{identita.2}
\lim_{N\rightarrow\infty}\omega( m_{N}(\boldsymbol{\sigma}))=\pm m(J,0)
\end{equation}
and then by applying to (\ref{identita.2}) the same procedure as shown above. The result is still given
by formulas (\ref{J.inverso.curie}) and (\ref{h.inverso.curie}) which conclude the analytical treatment of the inverse problem.

For what it concerns the statistical part one has to provide an evaluation of the finite size average magnetization
$\omega( m_{N}(\boldsymbol{\sigma}))$ and susceptibility $\chi_{N}$ from the empirical data. We use the maximum likelihood estimation procedure. This method identifies the free parameters 
within a distribution by requiring that their value
maximize the probability to obtain the given sample, under the condition that the sample is made of 
independent and identically distributed realizations of the random variables.

Given a  sample made of $M$ independent spin configurations $\boldsymbol{\sigma}^{(1)},\dots,\boldsymbol{\sigma}^{(M)}$ all distributed according to the Boltzmann-Gibbs measure (\ref{BG.curie}),  
the maximum likelihood \cite{Fi, J} function is defined by 
\begin{equation*}
 L(J,h)=P_{N,J,h}\Big\{\boldsymbol{\sigma}^{(1)},\dots,\boldsymbol{\sigma}^{(M)}\Big\}
\end{equation*}
which, using the independence, can be rewritten as
\begin{align*}
L(J,h)&=\prod_{i=1}^{M}P_{N,J,h}\Big\{\boldsymbol{\sigma}^{(i)}\Big\}\nonumber\\
&=\prod_{i=1}^{M}\frac{\exp(- H_{N}(\boldsymbol{\sigma}^{(i)}))}{\sum_{\boldsymbol{\sigma}\in\Omega_{N}}\exp(- H_{N}(\boldsymbol{\sigma}))}.
\end{align*}
To maximize the function $L(J,h)$ we should compute the derivative of a product. Since a function and its 
logarithm reach the maximum in the same point, we consider the logarithm of the maximum likelihood function
\begin{equation*}
\ln L(J,h)=\sum_{i=1}^{M}\bigg(-H_{N}(\boldsymbol{\sigma}^{(i)})-\ln\sum_{\boldsymbol{\sigma}\in\Omega_{N}}\exp(- H_{N}(\boldsymbol{\sigma}))\bigg).
\end{equation*}
The derivatives with respect to $h$ and $J$ of this function
\begin{align*}
\frac{\partial\ln L(J,h)}{\partial h}&=N\sum_{i=1}^{M}\bigg(m_{N}(\boldsymbol{\sigma}^{(i)})-\omega(m_{N}(\boldsymbol{\sigma}))\bigg)\\\nonumber
\frac{\partial\ln L(J,h)}{\partial J}&=\frac{N}{2}\sum_{i=1}^{M}\bigg(m_{N}^{2}(\boldsymbol{\sigma}^{(i)})-\omega(m_{N}^{2}(\boldsymbol{\sigma}))\bigg)
\end{align*}
vanish for
\begin{equation}\label{risultati.verosimiglianza}
\begin{cases}
\omega(m_{N}(\boldsymbol{\sigma}))=\dfrac{1}{M}\sum\limits_{i=1}^{M}m_{N}(\boldsymbol{\sigma}^{(i)})
\\ \\
\omega(m_{N}^{2}(\boldsymbol{\sigma})) =\dfrac{1}{M}\sum\limits_{i=1}^{M}m_{N}^{2}(\boldsymbol{\sigma}^{(i)}).
\end{cases}
\end{equation}
Therefore, the function $L(J,h)$ reaches its maximum when the first and second momentum of the magnetization are calculated from the data according to (\ref{risultati.verosimiglianza}). The inverse problem is finally solved
by the composition of (\ref{risultati.verosimiglianza}) with (\ref{J.inverso.curie}) and (\ref{h.inverso.curie}). In particular, denoting by $m_{exp}$ and $\chi_{exp}$ respectively the average magnetization and the susceptibility computed from the sample 
\begin{equation}\label{mexpchiexp}
m_{exp}=\dfrac{1}{M}\sum\limits_{i=1}^{M}m_{N}(\boldsymbol{\sigma}^{(i)}) \quad\quad \chi_{exp}=N\left(\dfrac{1}{M}\sum\limits_{i=1}^{M}m_{N}^{2}(\boldsymbol{\sigma}^{(i)})-m_{exp}^{2}\right),
\end{equation}
the estimators of the model's free parameters are
\begin{align}\label{stimatori}
J_{exp} &=\frac{1}{1-m_{exp}^{2}}-\frac{1}{\chi_{exp}}\\
h_{exp} &=\tanh^{-1}(m_{exp})-J_{exp}m_{exp}.\label{stimatorih}
\end{align}

\section{Inverse problem for the multi-species model}
Since, within the applications we are interested in, we aim at generalizing the 
discrete choice theory \cite{McF}, we proceed toward the solution of the inverse
problem for the multi-species mean field model. 
Formally such a model is an extension of the Curie-Weiss model to systems composed of many interacting groups.
We consider a system of $N$ particles that can be divided into $k$ subsets $P_{1},\dots , P_{k}$ with $P_{l}\cap P_{s}=\emptyset$, for $l\neq s$ and sizes $|P_{l}|=N_{l}$, where $\sum_{l=1}^{k}N_{l}=N$. Particles interact 
with each other and with an external field according to the mean field Hamiltonian:
\begin{equation}\label{Hami.multi}
H_{N}(\boldsymbol{\sigma})=-\frac{1}{2N}\sum_{i,j=1}^{N}J_{ij}\sigma_{i}\sigma_{j}-\sum_{i=1}^{N}h_{i}\sigma_{i} \; .
\end{equation}
The $\sigma_{i}\in\{\pm 1\}$ represents the spin of the particle $i$, while $J_{ij}$ is the parameter that tunes the mutual 
interaction between the particle $i$ and the particle $j$ and takes values according to the following symmetric matrix:

\begin{displaymath}
         \begin{array}{ll}
                \\
                N_1 \left\{ \begin{array}{ll||}
                                      \\
                                   \end{array}  \right.
                                        \\
                N_2 \left\{ \begin{array}{ll||}
                                        \\
                                   \end{array}  \right.
                                          \\
                                         \\
                                         \\
                      
                  N_k \left\{ \begin{array}{ll||}
                     \\
            \\
                \\
                                  \end{array}  \right.
         \end{array}
          \!\!\!\!\!\!\!\!
         \begin{array}{ll||}
                \quad
                 \overbrace{\qquad }^{\textrm{$N_1$}}\,
                 \overbrace{\qquad }^{\textrm{$N_2$}}\qquad\quad\;
                 \overbrace{\qquad\qquad\quad }^{\textrm{$N_k$}}
                  \\
                 \left(\begin{array}{c|c|cc|ccc}
                               \mathbf{ J}_{11}  &  \mathbf{ J}_{12} & &\;\dots\; & &\;\;\mathbf{ J}_{1k}\;\;&
                                \\
                                 \hline
                              \mathbf{ J}_{12} & \mathbf{ J}_{22} & & & & &\\
                             \hline
                             & & & & & &\\
                             \vdots & & & & & &\\
                             \hline
                             & & & & & &\\
                             \mathbf{ J}_{1k} & \mathbf{ J}_{2k} & &\;\dots\; & &\;\;\mathbf{ J}_{kk}\;\; &\\
                             & & & & & &
                      \end{array}\right)
               \end{array}
\end{displaymath}\\
\noindent where each block $\mathbf{J}_{ls}$ has constant elements $J_{ls}$. For $l=s$, $\mathbf{J}_{ll}$ is a square matrix, whereas the matrix $\mathbf{ J}_{ls}$ is rectangular. We assume $J_{11}, J_{22},\dots , J_{kk}$ to be positive, whereas $J_{ls}$ with $l\neq s$ can be either positive or negative allowing for both ferromagnetic and antiferromagnetic interactions. The vector field takes also different values depending on the subset the particles belong to as specified by the following vector: 

\begin{displaymath}
         \begin{array}{ll}
                N_1 \left\{ \begin{array}{ll}
                                      \\
                                   \end{array}  \right.
                                        \\
                N_2 \left\{ \begin{array}{ll}
                                        \\
                                   \end{array}  \right.
                                          \\
                                         \\
                                         \\
                      
                  N_k \left\{ \begin{array}{ll}
                     \\
            \\
                \\
                                  \end{array}  \right.

           \!\!\!\!\!\!
    \end{array}
    \!\!\!\!\!\!
    \left(\begin{array}{ccc|c}
                \mathbf{h}_{1}
            \\
            \hline
            
            \mathbf{h}_{2}
            \\
            \hline
            \\
            \vdots
            \\
            \hline
            \\
            \mathbf{h}_{k}
            \\
            \\
        \end{array}\right)
\end{displaymath}\\
\noindent where each $\mathbf{h}_{l}$ is a vector of constant elements $h_{l}$.
Indicating with $m_{l}(\boldsymbol{\sigma})$ the magnetization of the group $P_{l}$, and with $\alpha_{l}=N_{l}/N$ the relative size of the set $P_{l}$, we may easily express the Hamiltonian (\ref{Hami.multi}) as:
\begin{align}\label{Hami.multi.2}
H_{N}(\boldsymbol{\sigma}) &=-N\Big(\frac{1}{2}\sum\limits_{l, s=1}^{k}\alpha_{l}\alpha_{s}J_{ls}m_{l}(\boldsymbol{\sigma})m_{s}(\boldsymbol{\sigma})+\sum\limits_{l=1}^{k}\alpha_{l}h_{l}m_{l}(\boldsymbol{\sigma})\Big)\nonumber\\
&=-N\Big(\frac{1}{2}\langle\mathbf{J}\mathbf{D}_{\boldsymbol{\alpha}}\mathbf{m}(\boldsymbol{\sigma}),\mathbf{D}_{\boldsymbol{\alpha}}\mathbf{m}(\boldsymbol{\sigma})\rangle +\langle \mathbf{h}, \mathbf{D}_{\boldsymbol{\alpha}}\mathbf{m}(\boldsymbol{\sigma})\rangle\Big)
\end{align}
where $\mathbf{m}(\boldsymbol{\sigma})=(m_{1}(\boldsymbol{\sigma}),\dots ,m_{k}(\boldsymbol{\sigma}))$, $\mathbf{D}_{\boldsymbol{\alpha}}=diag\{\alpha_{1},\dots,\alpha_{k}\}$, $\mathbf{h}=(h_{1},\dots , h_{k})$ and $\mathbf{J}$ is the reduced interaction matrix
\begin{equation*}
\mathbf{J}=\begin{pmatrix}
J_{11}  & J_{12} & \dots & J_{1k}\\
J_{12}  & J_{22} & \dots & J_{2k}\\
\vdots&\vdots&&\vdots \\
J_{1k}  & J_{2k} & \dots & J_{kk}
\end{pmatrix}.
\end{equation*}
The joint distribution of a spin configuration $\boldsymbol{\sigma}=(\sigma_{1},\dots ,\sigma_{N})$ is given by the Boltzmann-Gibbs measure $P_{N, \mathbf{J}, \mathbf{h}}$ related to the Hamiltonian (\ref{Hami.multi}), where again we consider the inverse temperature parameter $\beta$ absorbed within the model parameters $\mathbf{J}$ and $\mathbf{h}$.  
The well position of the model has been shown in \cite{GC}. In particular, in the thermodynamic limit the model is described by the following system of mean-field equations:
\begin{equation}\label{campomedio.multi}
\begin{cases}
m_{1}(\mathbf{J},\mathbf{h}) &\!\!\!\!= \tanh\Big(\sum\limits_{l=1}^{k}\;\alpha_{l}J_{1l}\;m_{l}(\mathbf{J},\mathbf{h})+h_{1}\Big) \\
m_{2}(\mathbf{J},\mathbf{h}) &\!\!\!\!=\tanh\Big(\sum\limits_{l=1}^{k}\;\alpha_{l}J_{2l}\;m_{l}(\mathbf{J},\mathbf{h})+h_{2}\Big)\\
\;\vdots\\
m_{k}(\mathbf{J},\mathbf{h}) &\!\!\!\!=\tanh\Big(\sum\limits_{l=1}^{k}\;\alpha_{l}J_{lk}\;m_{l}(\mathbf{J},\mathbf{h})+h_{k}\Big) \; .
\end{cases} 
\end{equation}

If the system (\ref{campomedio.multi}) admits a unique thermodynamically stable solution $\mathbf{m}(\mathbf{J},\mathbf{h})=(m_{1}(\mathbf{J},\mathbf{h}),\dots,m_{k}(\mathbf{J},\mathbf{h}))$, the following identities hold (see \cite{F}):
\begin{equation}\label{identita.multi}
\lim_{N\rightarrow\infty}\omega(m_{l}(\boldsymbol{\sigma}))=m_{l}(\mathbf{J},\mathbf{h})\quad l=1,\dots,k.
\end{equation}

By differentiating the identities (\ref{identita.multi}) with respect to $h_{s}$, $s=1,\dots,k$, we obtain
\begin{equation}\label{derivata.identita.multi}
 \lim_{N\rightarrow\infty}\frac{\partial }{\partial h_{s}}\omega( m_{l}(\boldsymbol{\sigma}))=\chi_{ls}\quad l,s=1,\dots,k
\end{equation}
where $\chi_{ls}$ are the elements of the susceptibility matrix of the model. In particular,
\begin{align*}\label{chi.multi.1}
\chi_{ls}=\frac{\partial m_{l}(\mathbf{J},\mathbf{h})}{\partial h_{s}}&=\frac{\partial}{\partial h_{s}}\Big(\tanh\Big(h_{l}+\sum\limits_{p=1}^{k}\alpha_{p}J_{lp}m_{p}(\mathbf{J},\mathbf{h})\Big)\Big)\nonumber\\
&=(1-m_{l}^{2}(\mathbf{J},\mathbf{h}))\Big(\delta_{ls}+\sum\limits_{p=1}^{k}\alpha_{p}J_{lp}\chi_{ps}\Big)
\end{align*}
where $\delta_{ls}$ denotes the delta of Dirac picked in $l=s$. Therefore, the susceptibility matrix $\boldsymbol{\chi}$ can be written as:
\begin{equation}\label{chi.matrice}
\boldsymbol{\chi}=\mathbf{P}(\mathbf{I}+\mathbf{J}\mathbf{D}_{\boldsymbol{\alpha}}\boldsymbol{\chi})
\end{equation}
 where $\mathbf{P}=diag\{1-m_{1}^{2}(\mathbf{J},\mathbf{h}),\dots,1-m_{k}^{2}(\mathbf{J},\mathbf{h})\}$ and $\mathbf{I}$ is the identity matrix.
Moreover, for each $l,s=1,\dots,k$
\begin{align}\label{chi.multi.2}
\frac{\partial}{\partial h_{s}}\omega(m_{l}(\boldsymbol{\sigma}))
&=\frac{\partial}{\partial h_{s}}\bigg(\dfrac{\sum_{\boldsymbol{\sigma}\in\Omega_{N}}m_{l}(\boldsymbol{\sigma})e^{-H_{N}(\boldsymbol{\sigma})}}{\sum_{\boldsymbol{\sigma}\in\Omega_{N}}e^{-H_{N}(\boldsymbol{\sigma})}}\bigg)\nonumber\\
&=N_{s}\Big(\omega( m_{l}(\boldsymbol{\sigma})m_{s}(\boldsymbol{\sigma}))-\omega( m_{l}(\boldsymbol{\sigma}))\omega( m_{s}(\boldsymbol{\sigma}))\Big).
\end{align}

By computing the elements of $\boldsymbol{\chi}$ according to (\ref{derivata.identita.multi}) and (\ref{chi.multi.2}), by (\ref{chi.matrice}) we get an expression of the reduced interaction matrix $\mathbf{J}$ related to the average value and the correlations of the magnetizations in the thermodynamic limit:
\begin{equation}\label{J.inverso.multi}
 \mathbf{J}=(\mathbf{P}^{-1}-\boldsymbol{\chi}^{-1})\mathbf{D}_{\boldsymbol{\alpha}}^{-1} \; ,
\end{equation}
see \cite{L}.
Once the matrix $\mathbf{J}$ is determined, the elements of the vector $\mathbf{h}=(h_{1},\dots,h_{k})$ are obtained by inverting the mean field equations (\ref{campomedio.multi})
\begin{equation}\label{h.inverso.multi}
h_{l}=\tanh^{-1}\left(\lim_{N\rightarrow\infty}\omega( m_{l}(\boldsymbol{\sigma}))\right)-\sum\limits_{s=1}^{k}\;\alpha_{s}J_{ls}\lim_{N\rightarrow\infty}\omega( m_{s}(\boldsymbol{\sigma}))\quad l=1,\dots,k.
\end{equation}

The previous formulas (\ref{J.inverso.multi}) and (\ref{h.inverso.multi}) represent the analytical solution
of the inverse problem when the system of mean-field equations (\ref{campomedio.multi}) has a unique 
stable solution. When there are more stable solutions, identities (\ref{identita.multi}) have to be handled with care. 
Similar identities are, in fact, locally fulfilled around each solution and the solution of the inverse problem is still 
possible by applying to them the same procedure described above. The estimators are again given by (\ref{J.inverso.multi}) and (\ref{h.inverso.multi}). Before proceeding with the statistical part of the inverse problem we notice that it is well 
posed since, in particular, the degrees of freedom of the problem are equal to $k(k+3)/2$.

Also in this case, starting from phenomenological data, we proceed with the help
of the maximum likelihood principle. Consider a sample of $M$ independent spin configurations $\boldsymbol{\sigma}^{(1)},\dots,\boldsymbol{\sigma}^{(M)}$ distributed according to the Boltzmann-Gibbs measure $P_{N,\mathbf{J},\mathbf{h}}$;
the maximum likelihood function related to the sample is
\begin{align}\label{verosimiglianza.multi}
L(\mathbf{J},\mathbf{h})&=P_{N,\mathbf{J},\mathbf{h}}\Big\{\boldsymbol{\sigma}^{(1)},\dots,\boldsymbol{\sigma}^{(M)}\Big\}\nonumber\\
&=\prod_{i=1}^{M}P_{N,\mathbf{J},\mathbf{h}}\Big\{\boldsymbol{\sigma}^{(i)}\Big\}\nonumber\\
&=\prod_{i=1}^{M}\frac{\exp(- H_{N}(\boldsymbol{\sigma}^{(i)}))}{\sum_{\boldsymbol{\sigma}\in\Omega_{N}}\exp(- H_{N}(\boldsymbol{\sigma}))}.
\end{align}
Differentiating the logarithm of the likelihood function (\ref{verosimiglianza.multi})
\begin{equation*}
\ln L(\mathbf{J},\mathbf{h})=\sum_{i=1}^{M}\bigg(-H_{N}(\boldsymbol{\sigma}^{(i)})-\ln \sum_{\boldsymbol{\sigma}\in\Omega_{N}}\exp(- H_{N}(\boldsymbol{\sigma}))\bigg)
\end{equation*}
with respect to $h_{l}$ and $J_{ls}$, $l,s=1,\dots,k$ we obtain
\begin{align*}
\frac{\partial\ln L(\mathbf{J},\mathbf{h})}{\partial h_{l}}&=N_{l}\sum_{i=1}^{M}\bigg(m_{l}(\boldsymbol{\sigma}^{(i)})-\omega(m_{l}(\boldsymbol{\sigma}))\bigg)\\\
\frac{\partial\ln L(\mathbf{J},\mathbf{h})}{\partial J_{ls}}&=\frac{N\alpha_{l}\alpha_{s}}{2}\sum_{i=1}^{M}\bigg(m_{l}(\boldsymbol{\sigma}^{(i)})m_{s}(\boldsymbol{\sigma}^{(i)})-\omega(m_{l}(\boldsymbol{\sigma})m_{s}(\boldsymbol{\sigma}))\bigg).
\end{align*}

These derivatives are equal to zero as the following equalities hold
\begin{align}\label{ultimo.cap.5}
\begin{cases}
\omega(m_{l}(\boldsymbol{\sigma}))=\dfrac{1}{M}\sum\limits_{i=1}^{M}m_{l}(\boldsymbol{\sigma}^{(i)})\quad l=1,\dots,k\\\\
\omega(m_{l}(\boldsymbol{\sigma})m_{s}(\boldsymbol{\sigma}))=\dfrac{1}{M}\sum\limits_{i=1}^{M}m_{l}(\boldsymbol{\sigma}^{(i)})m_{s}(\boldsymbol{\sigma}^{(i)})\quad l,s=1,\dots,k.
\end{cases}
\end{align}

Therefore, the inverse problem for the multi-species model is solved
by the composition of (\ref{ultimo.cap.5}) with (\ref{J.inverso.multi}) and (\ref{h.inverso.multi}). In particular, denoting by $m_{l\; exp}$ the average magnetization of each specie calculated from the data 
\begin{equation*}
m_{l\; exp}=\dfrac{1}{M}\sum\limits_{i=1}^{M}m_{l}(\boldsymbol{\sigma}^{(i)})\quad l=1,\dots,k
\end{equation*}
and defined the matrices $\mathbf{P}_{exp}=diag\{1-m_{1\; exp}^{2},\dots,1-m_{k\; exp}^{2}\}$ and $\boldsymbol{\chi}_{exp}$, whose elements are 
\begin{equation*}
\chi_{ls\; exp}=N_{s}\left(\dfrac{1}{M}\sum\limits_{i=1}^{M}m_{l}(\boldsymbol{\sigma}^{(i)})m_{s}(\boldsymbol{\sigma}^{(i)})-m_{l\; exp}m_{s\; exp}\right)\quad l,s=1,\dots,k
\end{equation*}
the model estimators are
\begin{align}\label{StimatoribipJ}
\mathbf{J}_{exp} &=(\mathbf{P}_{exp}^{-1}-\boldsymbol{\chi}_{exp}^{-1})\mathbf{D}_{\boldsymbol{\alpha}}^{-1}\\
h_{l\; exp} &=\tanh^{-1}(m_{l\; exp})-\sum\limits_{s=1}^{k}\;\alpha_{s}J_{ls\; exp}m_{s\; exp}\quad l=1,\dots,k.\label{Stimatoribiph}
\end{align}

\section{The inversion at finite volume and finite sample size}
When dealing with real data the elegant exactly solvable model has to be replaced by
its finite size version. This is reflected both in the number of particles $N$ and in the
number $M$ of independent configurations in the sample, available from the statistical set. It is therefore
important to see how the inversion formulas perform for different values of those quantities
at assigned values of the parameters. 
The Curie Weiss model and its generalized multi-species
version provide an ideal testing set not only because most of the applications concern mean
field models but also because their finite size solution can still be handled thanks to 
the observation that the magnetization spectrum has a probability distribution that
can be exactly computed.

In this section we present a numerical test of our inversion procedure, both for
the Curie-Weiss model and for its multi-species version. 
In both cases, for each choice of the size
$N$ of system and of the free parameter values ($J$ and $h$ for the Curie-Weiss model, $\mathbf{J}$ and $\mathbf{h}$ for its multi-species generalization), the data that we are going to use are extracted from a virtually exact simulation of the equilibrium distribution. This is possible thanks to the mean-field nature of the models
(\ref{Hami.curie}) and (\ref{Hami.multi.2}), which reduces the computation of corresponding equilibrium distribution to that of the weights of the $O(N)$ values of the magnetization.
In this way, from
$P_{N,J,h}$ for the Curie-Weiss model and from $P_{N,\mathbf{J},\mathbf{h}}$ for its multi-species generalization, we can compute the finite size average magnetization and
susceptibility and extract sequences of configurations.\par
\begin{figure}
\centering
\includegraphics[width=0.8\textwidth]{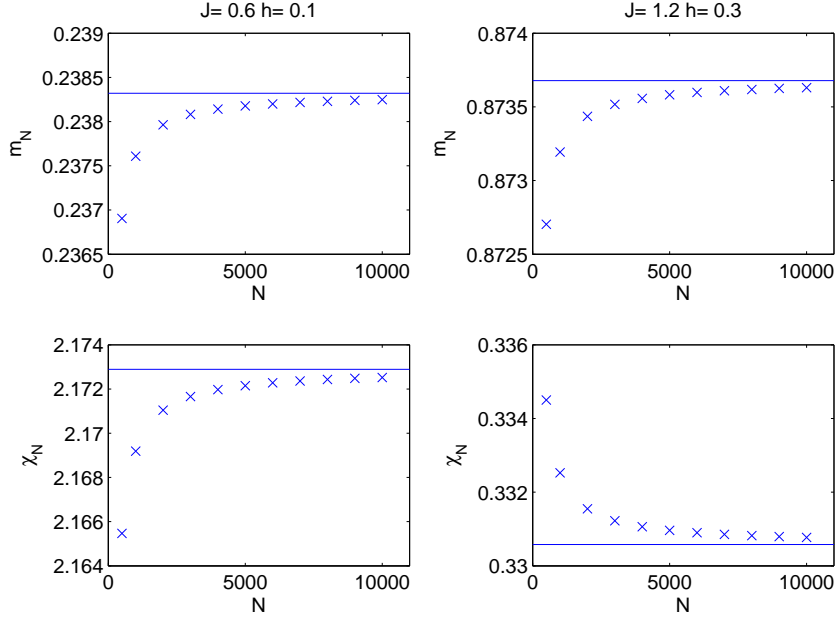}
\caption{\label{FigLimN} Finite size average magnetization $m_N$  (upper panels) and susceptibility $\chi_N$ (lower panels) as a function of $N$ for the Curie-Weiss model for $J=0.6$ and $h=0.1$ (left panels) and for $J=1.2$ and $h=0.3$ (right panel). The blue continuous lines represent the magnetization $m$ and the susceptibility $\chi$ in the thermodynamic limit.}
\end{figure}
\begin{figure}[h]
\centering
\includegraphics[width=0.8\textwidth]{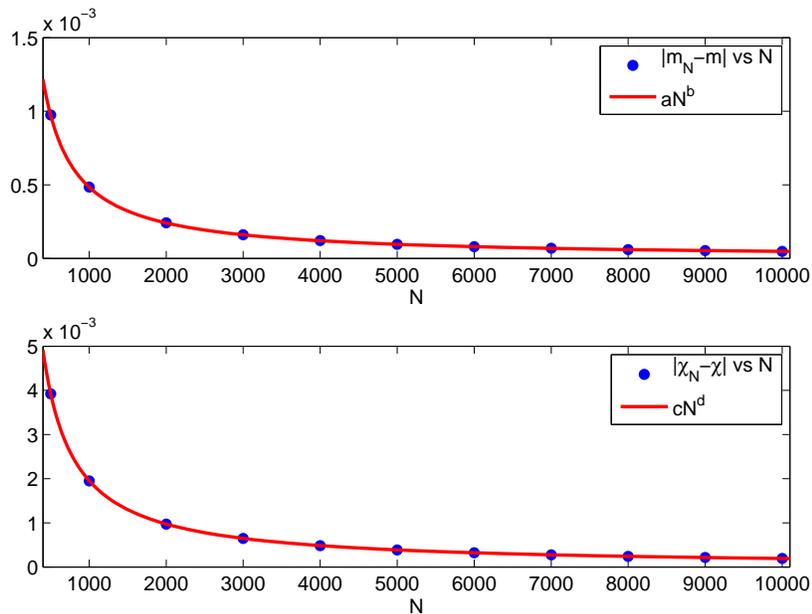}
\caption{\label{FigScalN} $J=1.2$ and $h=0.3$.  Upper panel:  $|m_N-m|$ as a function of $N$
together with the best fit $a N ^b$ for the data in the right upper panel of fig.\ref{FigLimN}. We obtain $a=0.5047 \pm 0.0037$
and $b=-1.006 \pm 0.002$ with a goodness of fit $R^2=1$.
Lower panel: $|\chi_N-\chi|$ as a function of $N$
together with the best fit $c N ^d$ for the data in the right lower panel of fig.\ref{FigLimN}. We obtain $c=2.037 \pm 0.019$
and $d=-1.006 \pm 0.002$ with a goodness of fit $R^2=1$.}
\end{figure}
\begin{figure}
\centering
\includegraphics[width=0.8\textwidth]{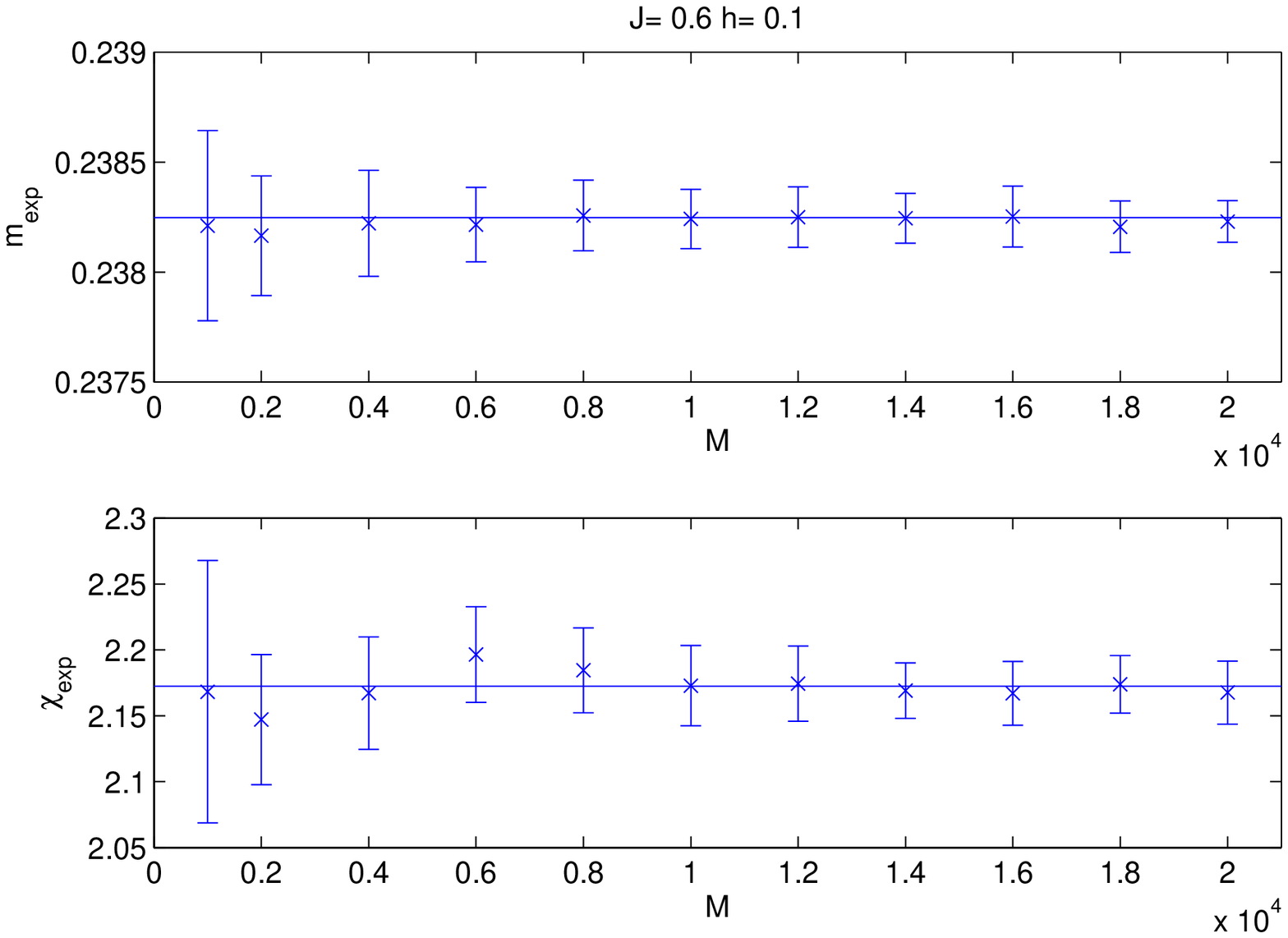}
\caption{\label{FigMsam}  $N=10000$, $J=0.6$ and $h=0.1$. Upper panel: Average magnetization $m_{exp}$  (blue crosses) as a function of $M$ (number of the configurations in the sample) together with statistical error bars over 20 $M$-sample for the Curie-Weiss model. The blue continuous line represents the finite size magnetization $m_N$ for $N=10000$.  Lower panel: Susceptibility $\chi_{exp}$ (blue crosses) as a function of $M$ (number of the configurations in the sample) together with statistical error bars over 20 $M$-sample. The blue continuous line represents the finite size susceptibility $\chi_N$ for $N=10000$}
\end{figure}

Although obvious, it is probably worth
remarking that the parameter estimation method involves two
approximations. The first one is in the inversion formulas
(\ref{J.inverso.curie}),(\ref{h.inverso.curie}),
(\ref{J.inverso.multi}) and (\ref{h.inverso.multi}),
that require the infinite volume limit; the second one is the
statistical error appearing in the evaluation of the averages and
correlations through the maximum-likelihood
estimators defined in (\ref{risultati.verosimiglianza}) and
(\ref{ultimo.cap.5}). In principle, the first approximation could
reduce strongly the scope of the method to systems with very large
number $N$ of individuals and it corresponds to estimating  the finite
size corrections of average magnetization and susceptibility. We don't go through
this issue, rather we illustrate it with some numerical example to support the choice
of the values of the parameters.
Indeed, figure \ref{FigLimN} shows the finite size average magnetization $\omega(m_N(\boldsymbol{\sigma}))$ and
susceptibility $N(\omega( m_{N}^{2}(\boldsymbol{\sigma}))-\omega^{2}( m_{N}(\boldsymbol{\sigma})))$ for the Curie-Weiss model at different $N$'s  
for $J=0.6$, $h=0.1$ and for $J=1.2$, $h=0.3$,
while the same quantities in the
thermodynamic limit $m(J,h)$ and $\chi$ are represented by the horizontal lines (to ease the notation, in the figure and in the following we denote the finite size quantities respectively by $m_N$ and $\chi_N$ and we omit the dependences of the equilibrium magnetization). The figure highlights the monotonic behavior of $m_{N}$ and $\chi_{N}$ as function of $N$. In particular, $m_{N}$ is monotonic increasing for each value of the interacting parameter $J$, while $\chi_{N}$ is monotonic increasing as $J<1$ and monotonic decreasing as $J>1$. We point out that the different behavior of the finite size susceptibility is very useful dealing with empirical data because it tells us if the system is above or under the interacting parameter critical value before to apply the inversion procedure. Note that, for $N\ge 5000$ we have optimal approximations
both for $m$ and $\chi$ in the thermodynamic limit. The
power-law fits in figure \ref{FigScalN} show evidence of the
$O(N^{-1})$-behavior of the finite size corrections  both for
magnetization and susceptibility, which entails the same  $O(N^{-1})$
error in the estimation of $J$ and $h$. Figure \ref{FigMsam} gives evidence of the dependence of the estimators $m_{exp}$ and $\chi_{exp}$, given by (\ref{mexpchiexp}),
on the choice of the number $M$
of the configurations of the sample that we use in the maximum-likelihood procedure.
To asses the statistical dependence of $m_{exp}$  and $\chi_{exp}$  on the sample 
$\{{\boldsymbol{\sigma}}^{(1)},\ldots, {\boldsymbol{\sigma}}^{(M)}\}$,  we computed their values over a set of $20$ independent instances of such samples. Thus from now on,  and without ambiguity, we use the subscript exp to denote both estimators and their statistical mean over the $20$ $M$-sample. We find numerical evidence that $M\ge 10000$ stabilizes the estimations. In particular, the standard deviation of both $m_{exp}$ and  $\chi_{exp}$ as a function of $M$ behaves as a power law $M^{-0.5}$ (as $J=0.6$ and $h=0.1$, the fit of the standard deviation of $m_{exp}$ is $aM^{-\alpha}$ with $\alpha=0.4933 \pm 0.06$, $a=0.013 \pm 0.006$, and goodness of fit $R^2=0.9696$, while those of the standard deviation of $\chi_{exp}$ is $bM^{-\beta}$ with $\beta=0.5175 \pm 0.098$, $b =3.269 \in (0.7388, 5.8)$ and $R^2=0.9343$). 

\begin{figure}
\centering
\includegraphics[width=0.8\textwidth]{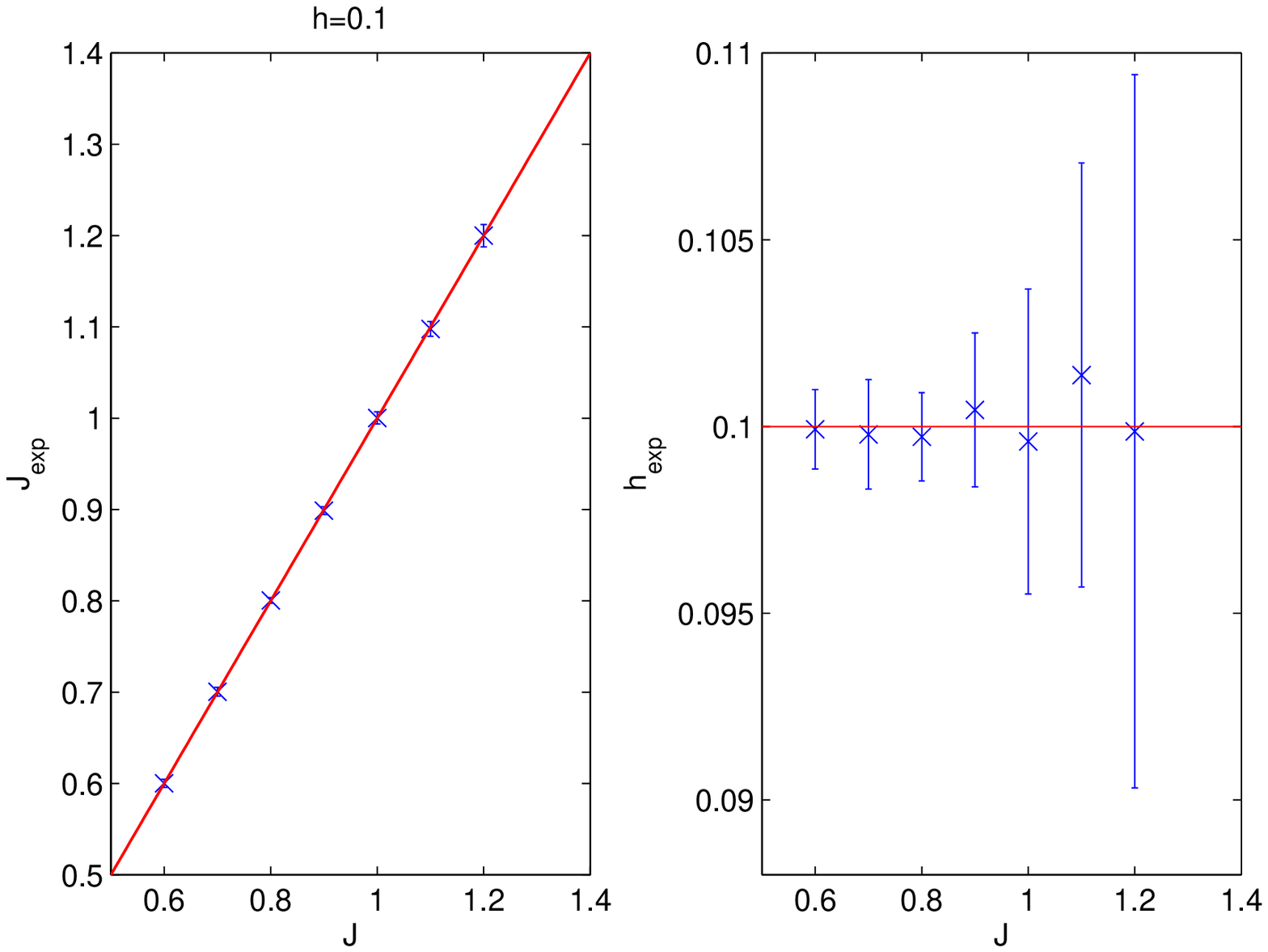}
\caption{\label{Fig1}  Left panel: $J_{exp}$ as a function of $J$ for $h=0.1$ (blue crosses). Error bars are standard deviations on $20$ different $M$-samples of the same simulation (see text for the details of the simulation). The red continous line represents $J_{exp}=J$.  Right panel: The value of $h_{exp}$ (blue crosses) calculated from (\ref{h.inverso.curie}) for the values of $J_{exp}$ in the left panel, as a function of $J$ together with the statistical error over $20$ $M$-samples. The horizontal line corresponds to the exact value of the magnetic field $h=0.1$}
\end{figure}

\begin{figure}
\centering
\includegraphics[width=0.8\textwidth]{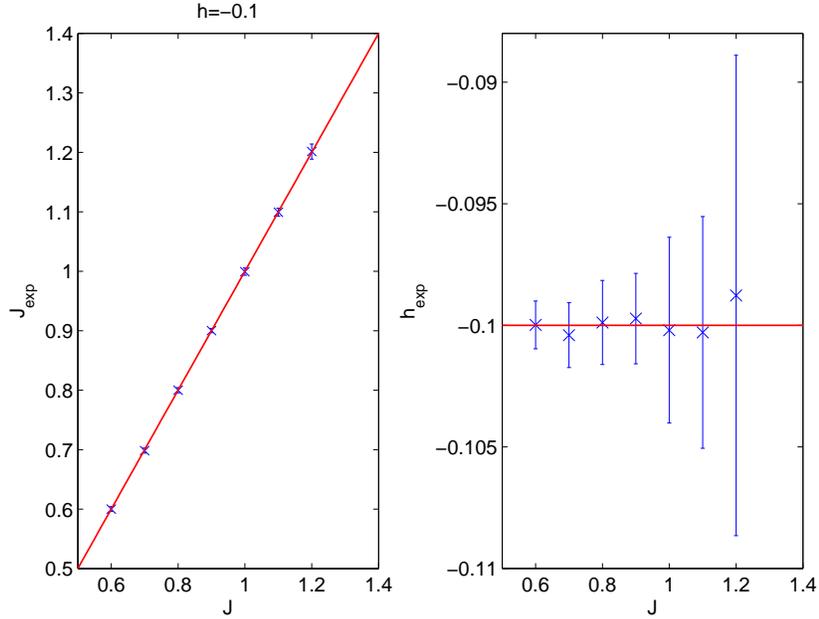}
\caption{\label{Fig2} Left panel: $J_{exp}$ as a function of $J$ for $h=-0.1$ (blue crosses). Error bars are standard deviations on $20$ different $M$-samples of the same simulation (see text for the details of the simulation). The red continuous line represents $J_{exp}=J$.  Right panel: The value of $h_{exp}$ (blue crosses) calculated from (\ref{h.inverso.curie}) for the values of $J_{exp}$ in the left panel, as a function of $J$ together with the statistical error over $20$ $M$-samples. The horizontal line corresponds to the exact value of the magnetic field $h=-0.1$}
\end{figure}

\begin{figure}
\centering
\includegraphics[width=0.8\textwidth]{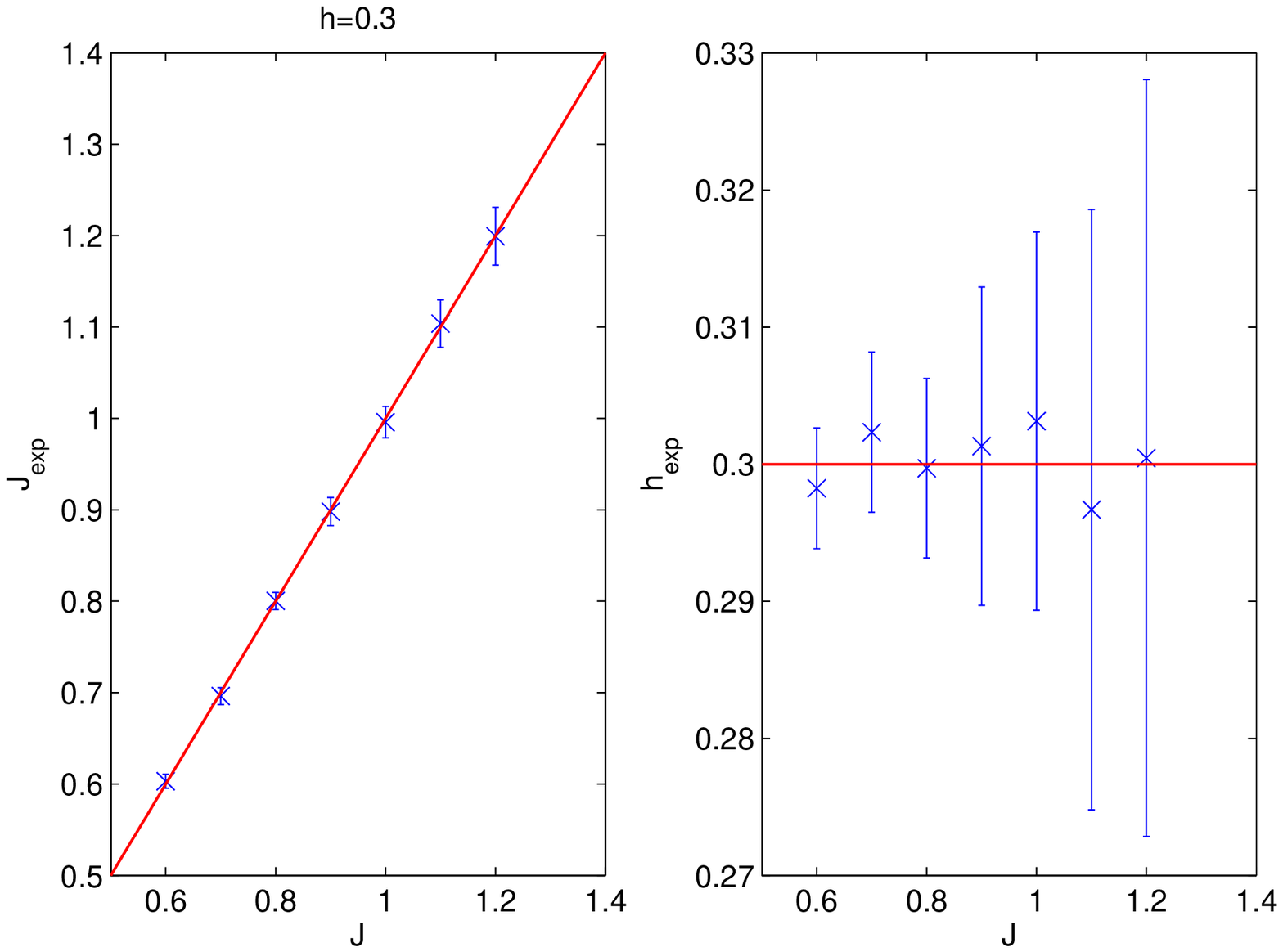}
\caption{\label{Fig3} Left panel: $J_{exp}$ as a function of $J$ for $h=0.3$ (blue crosses). Error bars are standard deviations on $20$ different $M$-samples of the same simulation (see text for the details of the simulation). The red continuous line represents $J_{exp}=J$.  Right panel: The value of $h_{exp}$ (blue crosses) calculated from (\ref{h.inverso.curie}) for the values of $J_{exp}$ in the left panel, as a function of $J$ together with the statistical error over $20$ $M$-samples. The horizontal line corresponds to the exact value of the magnetic field $h=0.3$}
\end{figure}

In order to test numerically the inversion procedure 
for the Curie-Weiss model, we consider a sample of $M=20000$ spin configurations
$\{\boldsymbol{\sigma}^{(i)}\}, i=1,\ldots,M$, where $\boldsymbol{\sigma}^{(i)}=(\sigma^{(i)}_1,\ldots,
\sigma^{(i)}_N)$ and $N=10000$. For a given couple of parameters $(J, h)$, we extract the sample of $M$ independent identically distributed spin configurations from the Boltzmann-Gibbs probability distribution function $P_{N, J,h}$. 
Given $(J,h)$, we consider $20$ $M$-sample and we solve the maximum likelihood model for each $M$-sample independently; then we average the inferred values $J_{exp}$ and $h_{exp}$ of the model parameters, given by (\ref{stimatori}) and (\ref{stimatorih}), over the 20 $M$-samples.
We consider $J\in [0.6,1.2]$ and $h\in [-0.3, 0.3]$. The obtained values for the case $h=0.1$ and $h=-0.1$ are shown in fig.\ref{Fig1} and in fig.\ref{Fig2}, where $J_{exp}$ and $h_{exp}$ are plotted as functions of $J$. Note that the 
inferred values of the parameters are in optimal agreement with the exact values (continuous lines in fig\ref{Fig1} and in fig.\ref{Fig2}).
The $J_{exp}$ and $h_{exp}$ values generated for $h=0.3$ are reported in fig.\ref{Fig3} to show that when the external field is enforced the prediction is quite good too: the points lie on the line of the exact agreement (red continuous lines in fig.\ref{Fig3}) even if the error bars are bigger than in previous cases.

\begin{figure}
\centering
\includegraphics[width=0.8\textwidth]{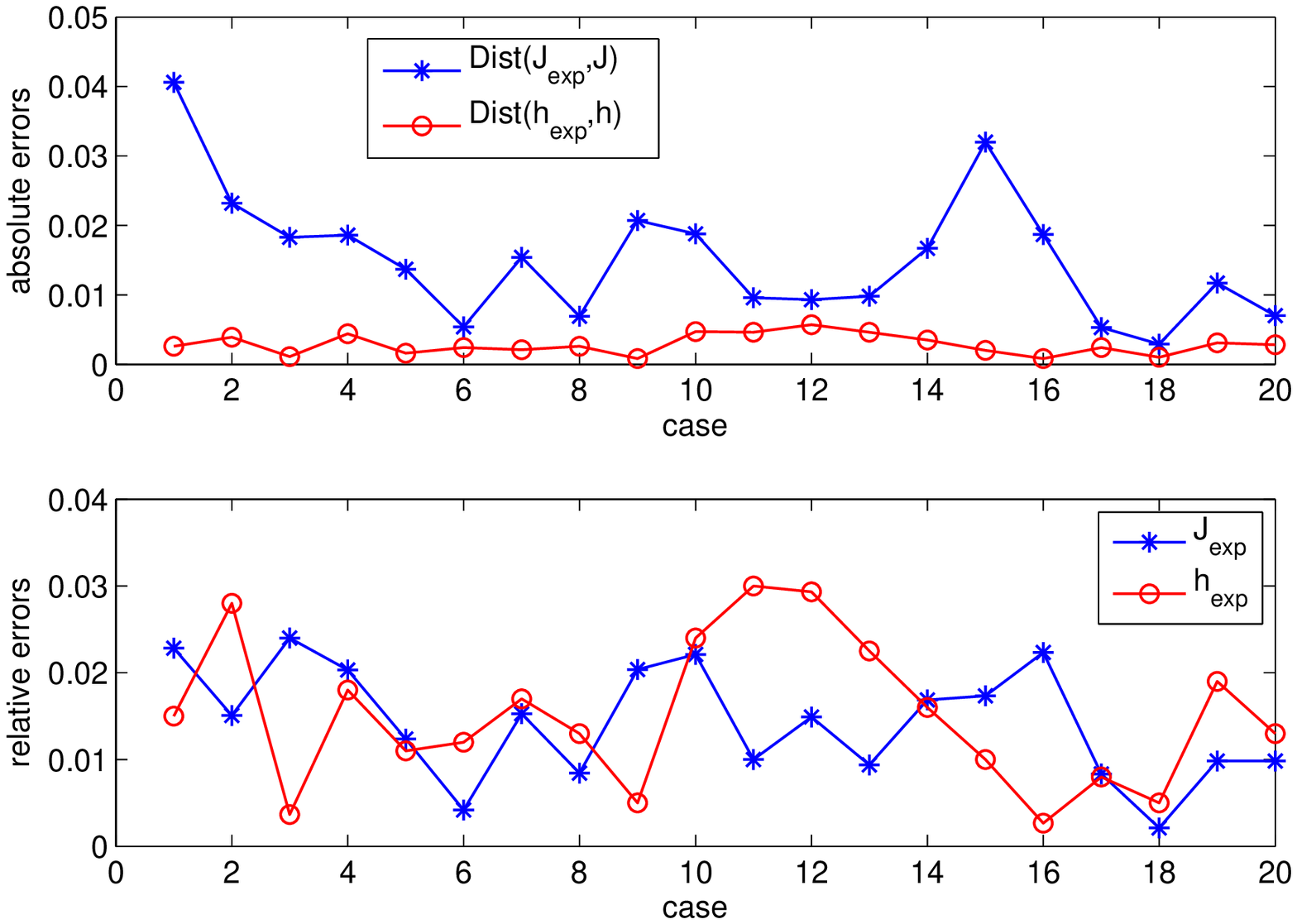}
\caption{Upper panel: Absolute errors in reconstructing $\mathbf{J}$ and $\mathbf{h}$. Distance between the reconstructed matrix $\mathbf{J}_{exp}$ and the initial data matrix $\mathbf{J}$ (blue stars) and distance between $\mathbf{h}_{exp}$ and the initial $\mathbf{h}$ (red circle) for $20$ different choices of parameters $\mathbf{J}$ and $\mathbf{h}$ (cases). The values of $\mathbf{J}_{exp}$ and $\mathbf{h}_{exp}$ are averaged across $20$ $M$-sample (see Section 4 in the text for the details of the simulations).
Lower panel: Relative errors in reconstructing $\mathbf{J}$ and $\mathbf{h}$. Maximum percentage errors for the reconstructed matrix $\mathbf{J}_{exp}$ (blue stars) and vector
$\mathbf{h}_{exp}$ (red circle) for the same $20$ multi-species cases considered in the upper panel.\label{Fig4} }
\end{figure}

As a test problem for the multi-species mean-field model we consider a system of $N=2000$ particles divided into $k=2$ equally populated subsets ($N_1=N_2=1000$) and a sample of $M=10000$ independent spin configurations. Starting from $20$ different couples of given values for the reduced interaction matrix 
\begin{equation}
\mathbf{J}=
\begin{pmatrix}
    J_{11}  &  J_{12} \\
    J_{12}  &  J_{22} 
\end{pmatrix}
\qquad J_{11},J_{22}\in[0.55,1.2],\quad  J_{12}\in[-0.6,1.1]
\end{equation}
and for the external vector field 
\begin{equation}
\mathbf{h}=
\begin{pmatrix}
    h_1 \\
    h_2
\end{pmatrix}
\qquad h_1,h_2\in [-0.3,0.3]
\end{equation}
we consider $20$ $M$-samples for each couple $(\mathbf{J},\mathbf{h})$ and we solve the maximum likelihood model for each one of them independently; then we average the inferred values $\mathbf{J}_{exp}$ and $\mathbf{h}_{exp}$ of the model parameters, given by (\ref{StimatoribipJ}) and (\ref{Stimatoribiph}), over the $20$ $M$-samples
(as in the one population model).
In fig.\ref{Fig4} the euclidean distances between  $\mathbf{J}_{exp}$ and the initial reduced interaction matrix $\mathbf{J}$ (blue stars) and between $\mathbf{h}_{exp}$ and the initial external vector field $\mathbf{h}$ (red circles) are shown for each of the $20$ choices of $\mathbf{J}$ and $\mathbf{h}$ (cases).  We observe that, as in the one-population model (Curie-Weiss model), the inverse problem procedure to infer the experimental values for the coupling matrix and for the external field produces results in very good agreement with the initial values. In order to have a quantitative measure of the goodness of this procedure, we focus on the two most representative cases.

If we consider the case $1$, for which the distance between the initial data matrix $\mathbf{J}$ and the inferred
matrix $\mathbf{J}_{exp}$ gives the  maximum value (see absolute errors in fig.\ref{Fig4}),  we have
\begin{equation}
\mathbf{J}=
\begin{pmatrix}
    1.2     &   0.98 \\
    0.98  &  0.8 
\end{pmatrix}
\quad\quad
\mathbf{J}_{exp}=
\begin{pmatrix}
1.173\pm 0.036	  & 0.993 \pm 0.028\\
0.993 \pm 0.028   &   0.794 \pm 0.040 
\end{pmatrix}
\end{equation}
and for the corresponding values of the external field

\begin{equation}
\mathbf{h}=
\begin{pmatrix}
    0.1 \\
    0.2
\end{pmatrix}
\quad\quad
\mathbf{h}_{exp}=
\begin{pmatrix}
0.102 \pm 0.012\\
0.198 \pm 0.011
\end{pmatrix}.
\end{equation}
The errors on each value of the matrix $\mathbf{J}_{exp}$ and the vector $\mathbf{h}_{exp}$ are standard deviations across $20$ different $M$-sample of the same $(\mathbf{J},\mathbf{h})$-simulation.

The case $18$ in fig.\ref{Fig4}, which gives the minimum value for the distance between $\mathbf{J}$ and $\mathbf{J}_{exp}$ corresponds to
\begin{equation}
\mathbf{J}=
\begin{pmatrix}
    0.6     &   -0.8 \\
    -0.8  &  0.9 
\end{pmatrix}
\quad\quad
\mathbf{J}_{exp}=
\begin{pmatrix}
0.601 \pm 0.022	  & -0.798 \pm 0.019 \\
-0.798 \pm 0.019 & 0.901 \pm 0.020
\end{pmatrix}
\end{equation}
and to the external field

\begin{equation}
\mathbf{h}=
\begin{pmatrix}
    -0.2 \\
    -0.3
\end{pmatrix}
\quad\quad
\mathbf{h}_{exp}=
\begin{pmatrix}
-0.201 \pm 0.005 \\
-0.300 \pm 0.005 
\end{pmatrix}.
\end{equation}
The errors on each value of the matrix $\mathbf{J}_{exp}$ and $\mathbf{h}_{exp}$ are the standard deviations across $20$ different $M$-samples of the same $(\mathbf{J},\mathbf{h})$-simulation.

\section{Conclusions and perspectives}

In this paper we have tested the robustness of the inversion method in a class of statistical
mechanics mean field models. The novelty of the results is both on the finite size behavior
of the exact solutions and on the quality of the inversion for finite number of samples. Our findings
show that with a modest investment on samples we are able to reconstruct the values of the parameters
within a few percentages. The relevance of the problem comes from the necessity to have a
fully tested method in the parameter evaluation from real data of socio-economic type, as started
from the seminal work of Brock and Durlauf. The nature of the investigated model belongs to those
without intrinsic randomness but we plan to extend a similar analysis to those cases with
random interactions like the Sherrington-Kirkpatrick model, and/or random network connections 
among agents, like in the diluted models.

\vspace{1cm}
\noindent {\bf Acknowledgments}: M. Fedele thanks the INdAM-COFUND Marie Curie fellowships for financial support.
The authors thank S. Galam and A. Vezzani for interesting discussions.


\begin{thebibliography}{CL}
\bibitem{BCGMSVW} W. Bialek, A. Cavagna, I. Giardina, T. Mora, E. Silvestri, M. Viale and A.M. Walczak, 
Statistical mechanics for natural flocks of birds,
Proceedings of the National Academy of Sciences, {\bf 109}, 4786-4791, (2012)

\bibitem{BD} W.A. Brock and S.N. Durlauf, 
Discrete choice with social interactions, 
Rev. Economic Studies, {\bf 68}, 235–260, (2001)

\bibitem{CG} P. Contucci and S. Ghirlanda,
Modeling Society with Statistical Mechanics: an Application to 
Cultural Contact and Immigration,
Quality and Quantity, {\bf 41}, 569-578, (2007)

\bibitem{E} R.S. Ellis, 
Entropy, large deviations, and statistical mechanics,
Springer (2005)

\bibitem{ENR} R.S. Ellis, C.M. Newman and J.S. Rosen,
Limit theorems for sums of dependent random variables occurring in statistical mechanics,
Probability Theory and Related Fields, {\bf 51}, 153-169, (1980)

\bibitem{F} M. Fedele, 
A mean field model for the collective behaviour of interacting multi-species particles: mathematical
results and application to the inverse problem, PhD thesis, University of Bologna, (2011)

\bibitem{Fi} R.A. Fisher,
Theory of statistical estimation,
Mathematical Proceedings of the Cambridge Philosophical Society, {\bf 22}, 700--725, (1925)

\bibitem{GC} I. Gallo and P. Contucci,
Bipartite mean field spin systems. Existence and solution,
Mathematical Physics Electronic Journal, {\bf 14}, 1-22, (2008)

\bibitem{Gr1} R. B. Griffiths, 
A proof that the free energy of a spin system is extensive, 
J. Math. Phys, {\bf 5}, 1215-1222, (1964)

\bibitem{Gr2} R. B. Griffiths, 
Correlation in Ising Ferromagnets, 
J. Math. Phys, {\bf 8}, 478-483, (1967)

\bibitem{Fa} F.E.Harrel, 
Regression modeling strategies: with applications to linear models, logistic regression and survival analysis,
Springer Series in Statistics, 2001

\bibitem{KS} D.G.Kelly and S. Sherman,
General Griffiths' Inequalities on Correlations in Ising Ferromagnets, 
J. Math. Phys. {\bf 9}, 466, (1968)

\bibitem{J} E.T. Janes,
Information theory and statistical mechanics,
Physical review, {\bf 106}, 620, (1957)

\bibitem{L} S. Loreti, Il problema inverso in meccanica statistica, Laurea Thesis, University of Bologna, (2010)

%\bibitem{McF1} D. McFadden,
%Modelling the choice of residential location,
%Institute of Transportation Studies, University of California, (1978)

\bibitem{McF} D. McFadden,
Economic Choices,
American Economic Review,  {\bf 91}, 351-378 (2001)

\bibitem{MB} Q. Michard and J.-P. Bouchaud,  
Theory of collective opinion shifts: from smooth trends to abrupt swings, 
European Physical Journal B {\bf 47},  151-159, (2005).

\bibitem{MC} R. Monasson and S. Cocco,
Fast inference of interactions in assemblies of stochastic integrate-and-fire neurons from spike recordings,
Journal of computational neuroscience, {\bf 31}, 199-227, (2011)

\bibitem{RTH} Y. Roudi, J. Tyrcha and J. Hertz,
Ising model for neural data: Model quality and approximate methods for extracting functional connectivity,
Physical Review E, {\bf 79}, 051915, (2009)

\bibitem{SM} V. Sessak and R. Monasson, 
Small-correlation expansions for the inverse Ising problem,
Journal of Physics A: Mathematical and Theoretical, {\bf 42}, 055001, (2009)

%\bibitem{T} T. Tanaka,
%Mean-field theory of Boltzmann machine learning,
%Physical Review E, {\bf 58}, 2302, (1998)

\bibitem{GGS} S. Galam, Y. Gefen and Y. Shapir, 
A mean behavior model for the process of strike, 
Mathematical Journal of Sociology {\bf 9}, 1-13, (1982)

\end{thebibliography}
\end{document}